\pdfoutput=1
\documentclass[runningheads,a4paper]{llncs}

\usepackage{amssymb}
\usepackage{amsmath}
\usepackage{graphicx}
\usepackage{color,soul}
\usepackage{multirow}
\usepackage{wrapfig}
\usepackage{booktabs}
\usepackage[misc]{ifsym} % For letter symbol
\usepackage[hidelinks]{hyperref}
\usepackage{microtype}

\usepackage{url}
\urldef{\mailsa}\path|{d.stoller, s.e.dixon}@qmul.ac.uk|
\urldef{\mailsb}\path|sewert@spotify.com|
\newcommand{\keywords}[1]{\par\addvspace\baselineskip
\noindent\keywordname\enspace\ignorespaces#1}

\newlength\myheight
\newlength\mydepth
\settototalheight\myheight{Xygp}
\newcommand*\inlinegraphics[2]{%
  \href{#2}{
  \settototalheight\myheight{Xygp}%
  \settodepth\mydepth{Xygp}%
  \raisebox{-\mydepth}{\includegraphics[height=\myheight]{#1}}%
  }
}

\selectfont

\begin{document}

\mainmatter  % start of an individual contribution

\title{Jointly Detecting and Separating Singing Voice: A Multi-Task Approach}

\titlerunning{Jointly Detecting and Separating Singing Voice: A Multi-Task Approach}

\author{Daniel Stoller\inst{1} \inlinegraphics{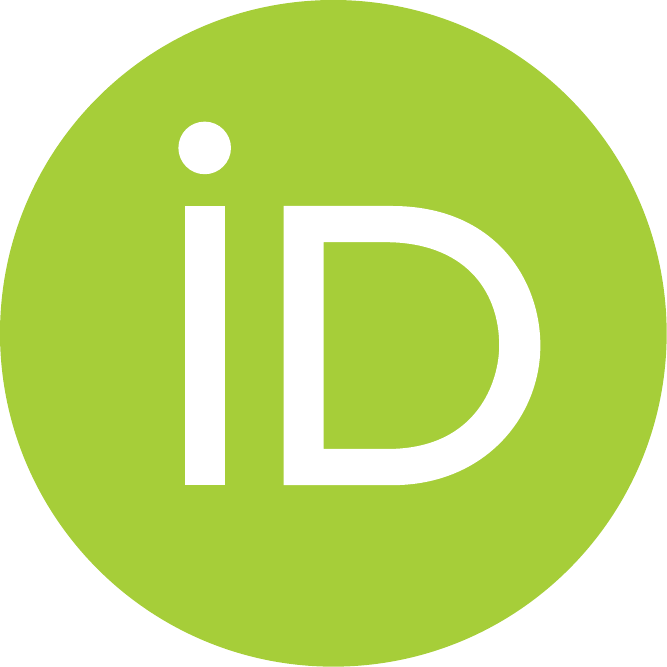}{https://orcid.org/0000-0002-8615-4144}\Letter %
\and Sebastian Ewert\inst{2} \inlinegraphics{orcid.pdf}{https://orcid.org/0000-0002-0718-0476} \footnote{Work was conducted at Queen Mary University of London.} \and Simon Dixon\inst{1} \inlinegraphics{orcid.pdf}{https://orcid.org/0000-0002-6098-481X}\Letter\ \footnote{This work was partially funded by EPSRC grants EP/L01632X/1 and EP/L019981/1.}
}
\authorrunning{Daniel Stoller \and Sebastian Ewert \and Simon Dixon}

\institute{Queen Mary University of London\\
London,
United Kingdom \\
\mailsa\\
%\url{http://www.springer.com/lncs}
\and
Spotify \\
London,
United Kingdom \\
\mailsb\\
%\url{http://www.springer.com/lncs}
}


\toctitle{Lecture Notes in Computer Science}
\tocauthor{Authors' Instructions}
\maketitle

\begin{abstract}
A main challenge in applying deep learning to music processing is the availability of training data.
One potential solution is Multi-task Learning, in which the model also learns to solve related auxiliary tasks on additional datasets to exploit their correlation.
While intuitive in principle, it can be challenging to identify related tasks and construct the model to optimally share information between tasks.
In this paper, we explore vocal activity detection as an additional task to stabilise and improve the performance of vocal separation.
Further, we identify problematic biases specific to each dataset that could limit the generalisation capability of separation and detection models, to which our proposed approach is robust.
Experiments show improved performance in separation as well as vocal detection compared to single-task baselines.
However, we find that the commonly used Signal-to-Distortion Ratio (SDR) metrics did not capture the improvement on non-vocal sections, indicating the need for improved evaluation methodologies.
\keywords{Singing voice separation, vocal activity detection, multi-task learning}
\end{abstract}

\section{Introduction and related work}

Separating the singing voice from the accompaniment in music recordings is a  challenging task, with the acoustical properties of the instruments involved and their interactions in a recording being highly complex.
Most current approaches train deep neural networks on multi-track recordings in a supervised fashion to estimate the individual sources from a given mixture input~\cite{Huang2014,Luo2017}.
While this approach often leads to considerable improvements over previous methods, it requires suitable input-output pairs from multi-track recordings. Unfortunately, publicly available datasets are often rather small on the order of a few hundred tracks.
This leads to overfitting and limits overall performance. 

\emph{Informed source separation} aims to circumvent this problem by providing additional information to the separation model, e.g.~the musical score~\cite{EwertS17_StructuredDropout_ICASSP}.
This way, the problem can be simplified, which often leads to improved results on small, annotated datasets.
On the other hand, such approaches can only be employed if suitable side information is indeed available, which is often not the case for musical scores.
In this paper, we thus focus on a more readily available and more easily created type of side information: vocal activity labels. 

A joint separation-classification model~\cite{Kong2017} was proposed for the more general problem of \textit{sound event detection} that employs a separation network whose output mask for each source is summarised with a mean or max operation to detect active sound events.
While similar to our approach, it is designed for weak labels and might be more sensitive to dataset biases when training with different separation and detection datasets due to its simple detection component.
Heittola \emph{et al.}~\cite{Heittola2011} use precise activity labels, but separation is used as a front-end for detection instead of performing joint estimation. Therefore, separation cannot be improved using mixtures with only activity labels.

To our knowledge, Chan \emph{et al.}~\cite{Chan2015} provide the only work combining \emph{singing voice separation (SVS)} in particular, with \emph{singing voice detection (SVD)}. Vocal activity labels are used to construct a mask, which forces the corresponding parts of the mixture spectrogram to be modelled individually in a method based on robust principal component analysis (RPCA).
For an increase in separation quality however, vocal activity labels are required during prediction.
The labels also have to be quite precise as a false negative label would force the vocal estimate to be zero for vocal sections.

Schl\"uter \cite{Schluter2016} focusses solely on SVD, but also shows that the resulting network can be used for detecting the location of the singing voice in the time-frequency domain.
This suggests it might be useful to integrate the information contained in the activity labels into separation models to improve their performance.
A related method was introduced by Ikemiya \emph{et al.}~\cite{Ikemiya2015}.
It produces a rough estimate of the vocals in a first step. After computing the fundamental frequency based on this estimate, the separation result is further refined. These two steps are repeated until convergence. We aim for a similar yet more integrated and joint estimation approach for the case of vocal activity labels.

\begin{figure}[t]
\includegraphics[width=\textwidth]{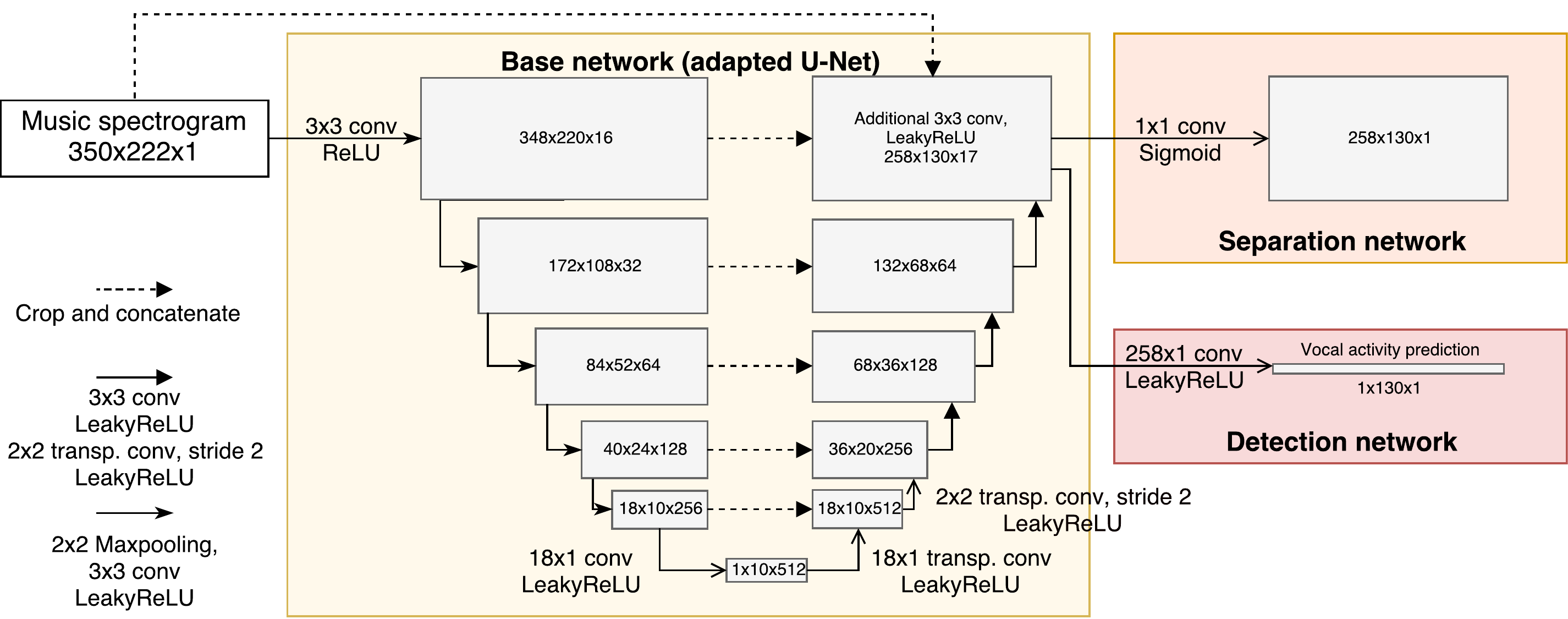}
\caption{Our multi-task model for jointly detecting and separating singing voice, given the spectrogram of a music piece as input. Tensor shapes are given in the order of frequency bins, time frames, and feature channels.}
\label{fig:model_diagram}
\end{figure}

Overall, vocal detection and separation are usually tackled as separate tasks despite their commonalities. Thus, a main goal in the this paper is to explore how such information can be exploited in training audio-only models that can jointly detect and separate vocals. 
First, we use a simple approach for diversifying the training dataset for an SVS model, and observe that its implicit assumption that all data sources are from the same distribution is violated due to a bias specific to each dataset.
Using a multi-task learning (MTL) approach, we then propose a model shown in Figure~\ref{fig:model_diagram} that performs SVS and SVD at the same time and can better account for such biases.
The model can be trained on multi-track recordings in combination with mixtures with vocal activity labels, and yields predictions on completely unlabelled mixtures.
By allowing the model to exploit correlations between the vocal activity labels and the source signals, performance is improved for both tasks compared to baseline models trained with single-task learning (STL).
While the overall improvement remained at a rather low level, we found the effect to be quite consistent -- despite the small size of the datasets involved and their respective biases.
We also found that the most commonly used evaluation metric~\cite{Vincent2006} is flawed in the sense that capturing improvements on non-vocal sections are not captured, and propose a simple ad-hoc solution.
As an additional contribution, we discuss the dataset biases we observed in some detail. Overall, based on these findings, we hypothesise that the joint prediction of source estimates along with side information such as musical scores in a multi-task setting could be a promising general direction for further research in music source separation. 

\section{Proposed approaches}

As a baseline system for SVS, we implemented a variant of the U-Net described in Section~\ref{sec:model} and shown in Figure~\ref{fig:model_diagram}. The approach is similar to~\cite{Stoller2018} and \cite{Jansson2017} and outputs a mask when given spectrogram magnitudes of a mixture excerpt.
During training, audio excerpts are randomly selected from the multi-track dataset, and converted to a log-normalised spectrogram representation.
The mean squared error (MSE) in spectral magnitudes between source estimates from the separator and the ground truth is used as a loss function.

\subsection{Initial approach to SVS: Using additional non-vocal sections}
\label{sec:approach_swapout}

Initially, we attempted to improve SVS performance by adding audio excerpts from instrumental sections of the SVD dataset to the SVS training set to increase its diversity:
Standard supervised training on a multi-track dataset entails randomly selecting audio excerpts from the tracks to generate batches of samples.
We changed this procedure so that when encountering an audio excerpt with silent vocals, it can be replaced with a randomly chosen non-vocal section from an additional music database with vocal activity labels.
The replacement occurs with a probability of $\frac{N}{N+M}$, with $N$ and $M$ being the size of the SVS and SVD dataset, respectively, to ensure non-vocal sections are effectively randomly sampled from both datasets.
To train from the additional non-vocal sections, we set their target accompaniment equal to the magnitudes of the respective mixture spectrogram, and all target magnitudes of the vocal spectrogram to zero.

The average MSE loss (see~\eqref{eq:mse}) on the test set obtained when training the same model with and without this replacement technique was used to test whether separation performance improves.
We performed the above training procedure with three different set-ups for the SVS and SVD dataset.

In the first experiment, we used the DSD100~\cite{Liutkus2017} dataset for SVS training, testing and evaluation, and RWC~\cite{Mauch2014} and Jamendo~\cite{Ramona2008} as the SVD dataset.
We also included a private collection of Dubstep, Hardstyle, Jazz, Classical and Trance music with 25 songs per genre.
We found that the performance decreased compared to purely supervised learning.
A first suspicion was that a bias in the test set might be responsible for inaccurate test performance measurements since only DSD100 is used (see section~\ref{sec:bias} for details).

To investigate this issue more closely, we conducted a second experiment and additionally included the MedleyDB~\cite{Bittner2014}, CCMixter~\cite{Liutkus2015} and iKala~\cite{Chan2015} SVS datasets in the validation and test sets.
Compared to the first experiment, the SVS training and test data is now less well matched, and the test performance gives a more accurate picture of generalisation capability.
Here performance increased considerably using our technique, strongly indicating that a bias in the SVS training data can be alleviated by including extra non-vocal sections.

Finally, we distributed the DSD100, MedleyDB, CCMixter and iKala datasets in equal proportions into training, validation and test set for a more realistic set-up in which all available multi-track data is used, but in this experiment, separation performance again decreased using our approach.

These results suggest that the individual datasets are subject to different biases in the data distribution space, to which our approach is sensitive since it assumes that all samples come from the same distribution.
These biases will be investigated in more detail in the next section.
Another shortcoming of our approach is that we cannot learn from the additional vocal sections using this method since we do not have the source audio available.

\subsection{Dataset bias for singing voice separation and detection}
\label{sec:bias}

\begin{figure}[t]
\includegraphics[width=1.0\textwidth]{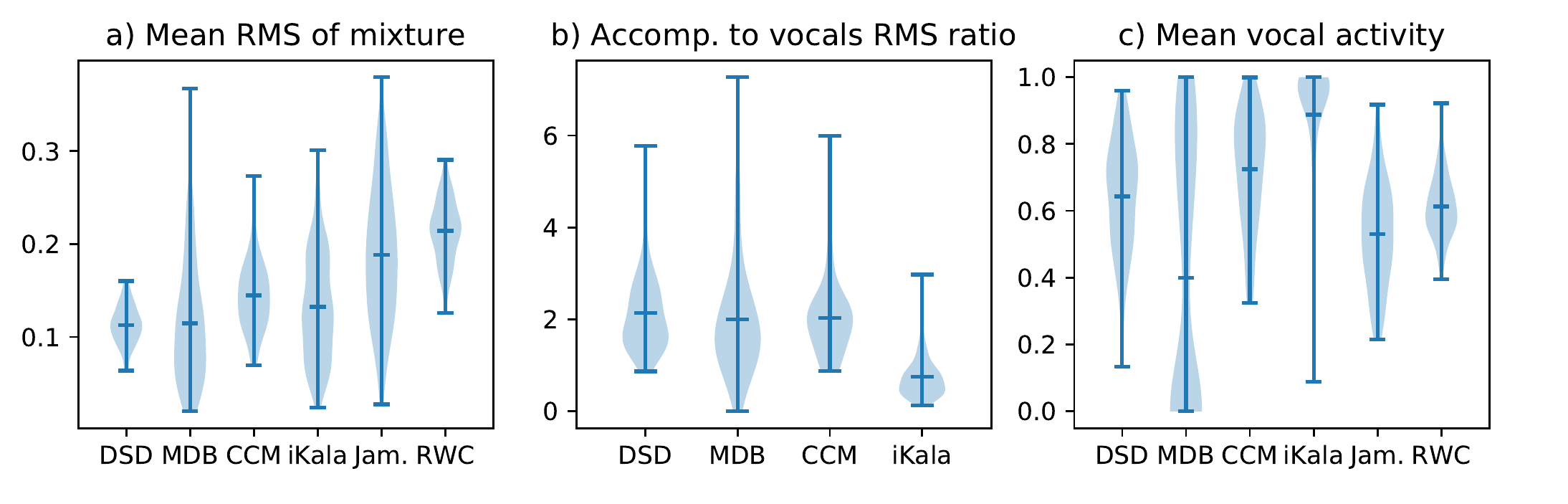}
\vspace{-0.7cm}
\caption{Distribution of values for different collections of tracks, for different properties. Outliers for MedleyDB in b) resulting from instrumental tracks have been excluded.}
\label{fig:bias_plot}
\vspace{-0.2cm}
\end{figure}

Since we are combining data from different sources, it is important to consider the impact of dataset bias on the performance of models trained on such combined data.
We hypothesised that datasets used for SVD and SVS are each uniquely biased, which can include properties such as the relative energy of the sources, overall energy levels and how often vocals occur on average.
We computed metrics for the above for the MedleyDB, DSD100, CCMixter, iKala, Jamendo and RWC datasets, as they are commonly used for SVD and SVS.
Vocals were considered active if the average absolute amplitude in a 10 ms window exceeded $5 \cdot 10^{-4}$.
Figure~\ref{fig:bias_plot} shows the distribution of these properties for each dataset, where metrics have been averaged song-wise.

Clear dataset bias manifests itself in the uneven distribution of values across datasets. For example, iKala contains relatively loud vocals and very few instrumental sections, and CCMixter has louder tracks than DSD100 with more vocals on average.
Additionally, even more dataset bias could be present in features which are more difficult to detect and quantify, such as timbre, language of the lyrics, music genre, recording conditions or the bleed level for multi-track recordings.
Therefore, it is very difficult to directly prevent models from overfitting to these biases. 
We would like to highlight this as a critical problem for the field of SVS and SVD, since many models are trained on a single dataset source and thus may not generalise nearly as well as the test scores indicate.

\subsection{Multi-task learning approach}

To mitigate problems due to dataset biases, we employ a multi-task learning (MTL) approach~\cite{Caruana1998} instead.
We augment the separation model with a component that predicts vocal activity based on a hidden layer of the separation model.
We train the combined model to output the source signals in the multi-track dataset and the vocal activity labels in the SVD dataset, respectively, with most parameters being shared for both tasks.

This approach has multiple benefits.
Firstly, predicting both outputs based on a shared hidden representation only assumes that the source output has some relationship with human-annotated vocal activity labels, but we do not define it explicitly.
For example, temporal inaccuracy in labels could mean that the beginnings of vocals are annotated as non-vocal. If we force the vocal output of the separator to be silent for all sections annotated as non-vocal, or use the approach from Section~\ref{sec:approach_swapout}, we give incorrect information to the separator. 
Secondly, a different dataset bias for each task can be accounted for by the model to some extent with its task-specific components. 
Thirdly, we exploit the information present in extra non-vocal and vocal sections.
Finally, the trained model can be used for both SVS and SVD.

For the SVS task, we use the MSE between the separator prediction $f_{\phi}(\mathbf{m})$ for a mixture excerpt $\mathbf{m}$ and the true sources $\mathbf{s}$ as the loss:
\begin{equation}
L_{\text{MSE}} = \mathbb{E}_{(\mathbf{m},\mathbf{s}) \sim p^1} \frac{1}{N}\ ||\mathbf{s} - f_{\phi}(\mathbf{m})||^2
\end{equation}

where $p^1$ represents the multi-track dataset distribution, which is approximated by a batch of samples, and $N$ denotes the dimensionality of the joint source vectors $\mathbf{s}$ and $f_{\phi}(\mathbf{m})$.
For output spectrograms with $T$ time frames, $F$ frequency bins and $K$ sources, $N = T \cdot F \cdot K$.

For the SVD task, we use the binary cross-entropy at each time frame of the spectrogram excerpt, averaged over time and over data points:
\begin{equation}
L_{\text{CE}} = \mathbb{E}_{(\mathbf{m}, \mathbf{o}) \sim p^2} \frac{1}{T} \sum_{t=1}^T \log p^t_{\phi}(o_t | \mathbf{m})
\end{equation}
where $p^t_{\phi}$ denotes the probability of the vocal state the model assigns to time frame~$t$ of the audio excerpt with a total of $T$ frames, and
$p^2$ describes the SVD dataset distribution whose samples contain a binary vector $\mathbf{o}$ with a vocal activity label $o_t$ at each spectral frame $t$ of the source output spectrogram.

For our MTL model, we combine the two above losses using a simple weighting scheme:
\begin{equation}
L_{\text{MTL}} = \alpha L_{\text{MSE}} + (1 - \alpha) L_{\text{CE}}.
\label{eq:mse}
\end{equation}
We set $\alpha = 0.9$ so that experimentally the losses are approximately on the same scale during training.
Although an optimisation of this hyper-parameter might improve results further, it is omitted here due to computational cost.
We also experimented with a loss function derived from a Maximum Likelihood objective
(see ancillary file) % For arXiv submission
%\footnote{Test} % For camera-ready submission
, but did not obtain better performance.

\section{Evaluation}

Next, we describe the experimental evaluation procedure for our MTL approach.

\subsection{Datasets}

For the SVS dataset, we use DSD100 with 50 songs each for training and testing, according to the predefined split.
We use the Jamendo dataset for SVD, since it predominantly contains Western Pop and Rock music, similarly to DSD100, to avoid a large dataset bias.
Jamendo's validation and test partitions comprising 30 songs are used for testing, leaving 60 songs for training.
This set-up is intended as a proof of concept of the MTL approach~--~in this setting even slight improvements are promising, since vocal activity labels do not directly yield information on vocal structure, and should translate to larger improvements given larger SVS and particularly SVD datasets.

\subsection{Model architecture and preprocessing}
\label{sec:model}

The audio input is converted to mono and down-sampled to $22050$ Hz to reduce dimensionality, before the magnitude spectrogram is computed from a 512-point FFT with 50\% overlap, and normalised by ${x \rightarrow }\log(1+x)$. Excerpts comprised of 222 time frames each are used as input to our model shown in Figure~\ref{fig:model_diagram}, which consists of a base network that branches off into a separation and a detection network.

The \textbf{base network} closely follows our previous implementation~\cite{Stoller2018} of the U-Net~\cite{Jansson2017}.
The output of an initial $3 \times 3$ convolution with $16$ filters and ReLU non-linearity is fed to a down-sampling block consisting of max-pooling with size and stride two followed by a $3 \times 3$ convolution with $32$ filters. The down-sampling block is applied three more times, each time doubling the number of filters, finally yielding a $18 \times 10 \times 256$ feature map.
We then use a 1D convolution with filter size $18 \times 1$ before applying the respectively transposed convolution, and concatenate it with the original $18 \times 10 \times 256$ feature map to capture frequency relationships.
In the following up-sampling block, a $2 \times 2$ transposed convolution with $128$ filters is applied, and the output concatenated with the output of the down-sampling block at the same network depth after centre-cropping it.
Lastly, a $3 \times 3$ convolution with $128$ filters is applied.
After applying this up-sampling block another three times, each time with half as many filters for the convolutions, the resulting $258 \times 130 \times 16$ feature map is concatenated with the centre-cropped input.
The resulting features are input to the SVS well as the SVD sub-network.

The output size is smaller than the input size since we use ``valid" convolutions that do not employ implicit zero-padding.
Therefore, the mixture naturally provides additional temporal context processed during convolution, and its magnitudes are zero-padded in frequency so that the separator output has the correct number of frequency bins.
Unless otherwise stated, Leaky ReLU is used after all convolutions as non-linearities to allow for better gradient flow.

In the \textbf{SVS network}, the feature map from the base architecture is transformed into a filtering mask, which is multiplied point-wise with the original mixture spectrogram magnitudes to yield the source estimates.
To generate the source audio, we use an inverse STFT using the mixture's phase, and apply $10$ iterations of the Griffin-Lim algorithm~\cite{Griffin1984} to further refine the phase.

The \textbf{SVD network} takes the final feature map from the base architecture and applies a single $F \times 1$ filter, where $F$ is the number of frequency bins, to reduce the time-frequency feature map to a single scalar for each time step.
Application of a sigmoid non-linearity yields the probability of the presence of singing voice at each time step.

\subsection{Experimental set-up and metrics}

To identify the impact of our proposed approach in comparison to solving separation and detection separately, we train and evaluate our network solely for either SVS or SVD, before comparing to training with the multi-task loss.

Model performance is evaluated on the test dataset every 1000 iterations and the model with the best performance is selected. Training is stopped after 10,000 iterations without performance improvement.
For SVD, we use the \textit{area under the receiver operating characteristic (AU-ROC)} to evaluate performance.
For separation, we use the MSE training objective from~\eqref{eq:mse} in the normalised magnitude space, as well as the track-wise SDR, SIR, and SAR metrics~\cite{Vincent2006} on the audio signals.
We select two MTL models with the best AU-ROC or MSE value, respectively, since best performance is reached at different training stages.

\subsection{Results}

Table~\ref{tab:model_results} shows a performance comparison of the considered models.
For both SVD and SVS, we achieve a slight improvement in both AU-ROC and MSE performance metrics using our model variants.
This is promising since the SVD dataset is small and vocal activity labels are less informative training targets than the vocals themselves.
Therefore, larger datasets could be used in future work to obtain larger performance increases.

\begin{table}[t]
\centering
\begin{tabular}{ccccccccccc}
\toprule
& & \multicolumn{9}{c}{Metric} \\
\cmidrule{3-11}
& & & & & \multicolumn{3}{c}{Vocals} & \multicolumn{3}{c}{Accompaniment} \\
\cmidrule(l{2pt}r{2pt}){6-8} \cmidrule(l{2pt}r{2pt}){9-11}
& & AU-ROC & MSE & Non-voc. RMS\ & SDR & SIR & SAR & SDR & SIR & SAR \\
\midrule
& SVD & 0.9239 & - & - & - & - & - & - & - & - \\
Model & SVS & - & 0.01865 & 0.0194 & 2.83 & 5.27 & \textbf{6.88} & 6.71 & \textbf{14.75} & 13.25 \\
& Ours & \textbf{0.9250} & \textbf{0.01755} & \textbf{0.0155} & 2.86 & \textbf{5.56} & 6.23 & 6.69 & 13.24 & \textbf{14.11} \\
\bottomrule
\end{tabular}
\caption{Performance comparison between SVS and SVD baseline and our approach. Results significantly better than the comparison model ($p < 0.05$) in bold. Significance of the AU-ROC difference determined with binary labels from all time frames as samples~\cite{Delong1988}. A paired Wilcoxon signed-rank test was used for all other metrics.}
\label{tab:model_results}
\vspace{-0.4cm}
\end{table}

While the MSE on the normalised spectrogram magnitudes improves by about 6\%, the mean SDR for vocals and accompaniment does not change significantly.
To find the cause, we analyse the employed implementation for SDR computation on the DSD100 dataset~\footnote{\url{https://github.com/faroit/dsd100mat}} also used in the SiSec source separation evaluation campaign~\cite{Liutkus2017}.
Tracks are partitioned into excerpts of 30s duration, using 15s of overlap, for which a local SDR value is computed. The final SDR is the average of the local SDR values.
However, for excerpts where at least one source is completely silent, the SDR has an undefined value of $\log(0)$ and is excluded from the final SDR average, so that the model's performance in these sections is ignored.
This is the case for 79 of 736 excerpts due to non-vocal sections and is thus a practically relevant flaw of the evaluation metric.

More sophisticated methods such as~\cite{Vincent2012} take audio perception more explicitly into account, but presumably suffer from the same issue with silent sources, as similar computations are used there as well.
As an ad-hoc solution, we propose computing the source estimate's energy or ideally loudness for silent sections of the source ground truth as a simple work\-around and report it in addition to other metrics.
Finding a consistent and perceptually accurate evaluation metric is thus an important unsolved problem, and listening tests arguably remain important to accurately assess separation quality.

A lower average MSE combined with a stagnating SDR suggests that our model improves especially on these non-vocal sections excluded from the SDR, potentially because negative vocal activity labels allow the separator to detect many different instruments as not being vocals.
To test this more explicitly, we take the vocal estimates of the baseline and our model and compute the average RMS of the 79 excerpts excluded from SDR computation, as well as the average output over whole songs in the DSD100 dataset.
We find that our model has less energy in its vocal output compared to the baseline, but also in the non-vocal sections (see Table~\ref{tab:model_results}).
This demonstrates that our model performs better on non-vocal sections and about equally on vocal sections due to a similar SDR.

\section{Conclusions}

We demonstrated that jointly solving the task of singing voice detection and singing voice separation can improve performance in both tasks and alleviates the issue of dataset scarcity.
Furthermore, we found biases specific to each dataset that could prevent source separation and detection models from generalising properly to unseen data.
Finally, we discuss a major flaw in the most popular evaluation metric for source separation~\cite{Vincent2006} related to the performance measurement in silent sections.

Therefore, further research into improved, perceptually relevant metrics is a definite need. As a workaround, we propose additionally measuring and reporting the loudness of the model's source estimates for sections where the respective source is silent.
Our multi-task approach could be generalised and applied to mixtures with pitch curve or phoneme annotations of the singing voice, or even to whole transcriptions of musical sources (see~\cite{Benetos2013}).
Performance increases can be expected to be larger especially for the latter case as correct predictions on one task greatly simplify solving the other one.

\vspace{0.4cm}
\noindent \textbf{Acknowledgements:} We thank Emmanouil Benetos for the useful comments and feedback, as well as Mi Tian for references on related literature.

\bibliographystyle{splncs03}
\bibliography{refs}

\end{document}